\documentclass[aps,prb,reprint,superscriptaddress]{revtex4-2}

\usepackage{graphicx}
\usepackage{grffile}
\usepackage{siunitx}
\usepackage{tabularx}
\usepackage[justification=raggedright, labelformat=simple]{caption}
\usepackage[singlelinecheck=off, justification=raggedright, position=top]{subcaption}
\usepackage[english]{babel}
\usepackage{physics}
\usepackage{cleveref}
\usepackage{todonotes}
\usepackage{csvsimple}
\usepackage{bm}

\usepackage{tikz}
\usetikzlibrary{arrows}
\usetikzlibrary{arrows.meta}
\usetikzlibrary{calc}
\usepackage{lipsum}

\begin{document}
\title{Spin and orbital Hall currents detected via current-induced magneto-optical Kerr effect in V and Pt}

\author{Yukihiro Marui}
\affiliation{Department of Physics, The University of Tokyo, Tokyo, 113-0033, Japan}
\affiliation{Research Institute of Electrical Communication, Tohoku University, Sendai, 980-8577, Japan}

\author{Masashi Kawaguchi}
\affiliation{Department of Physics, The University of Tokyo, Tokyo, 113-0033, Japan}

\author{Satoshi Sumi}
\affiliation{Toyota Technological Institute, Nagoya, 468-8511, Japan}

\author{Hiroyuki Awano}
\affiliation{Toyota Technological Institute, Nagoya, 468-8511, Japan}

\author{Kohji Nakamura}
\affiliation{Graduate School of Engineering, Mie University, Tsu, 514-8507, Japan}

\author{Masamitsu Hayashi}
\affiliation{Department of Physics, The University of Tokyo, Tokyo, 113-0033, Japan}
\affiliation{Trans-scale quantum science institute, The University of Tokyo, Tokyo, 113-0033, Japan}

\date{\today}

\begin{abstract}
We have studied the film thickness dependence of the current-induced magneto-optical Kerr effect in Pt and V thin films. The Kerr signal for Pt shows little dependence on the thickness in the range studied (20-80 nm). In contrast, the signal for V increases with increasing thickness and saturates at a thickness near 100 nm to a value significantly larger than that of Pt. These experimental results are accounted for assuming that spin and orbital Hall effects are responsible for the Kerr signal. We show that the Kerr signal is proportional to the product of the dc spin (orbital) Hall conductivity and the energy derivative of the ac spin (orbital) Hall conductivity. Contributions from the spin and orbital Hall effects mostly add up for V whereas they cancel out for Pt. Assuming that the orbital Hall conductivity matches that predicted from first-principles calculations, the thickness dependence of the Kerr signal suggests that the orbital diffusion length of V is considerably smaller compared to its spin diffusion length.
\end{abstract}
\maketitle


\section{Introduction}
Efficient generation of spin current is one of the key challenges in developing random access and storage class memory technologies\cite{parkin2015nnano,dieny2020nelec}. The spin Hall effect\cite{dyakonov1971jetp,murakami2003science} of the 5$d$ transition metals is considered to be, by far, the most viable approach in generating spin current using materials that are compatible with semiconductor manufacturing processes\cite{hoffmann2013ieee}. The efficiency to generate spin current is often represented by the spin Hall angle of the host material. The spin Hall angle is proportional to the product of the spin Hall conductivity and the resistivity. With regard to technological applications, it is preferable to reduce the resistivity to limit power consumption. Significant effort has thus been put into finding/developing materials with large spin Hall conductivity.

It has been shown that, in the 5$d$ transition metals, the spin Hall conductivity is determined by its band structure and/or spin-dependent scattering\cite{sinova2015rmp}. The latter, often referred to as the extrinsic spin Hall effect, can enhance the spin Hall conductivity if the appropriate element is doped as a scattering center\cite{niimi2011prl,sagasta2018prb}. The former, i.e., the intrinsic spin Hall effect, is determined by the electronic structure of the host material\cite{guo2008prl}. Studies have shown that the intrinsic spin Hall conductivity of transition metals follows a Hund's-rule-like scaling and is proportional to the $\bm{L} \cdot \bm{S}$ spin orbit coupling, where $\bm{S}$ and $\bm{L}$ are the total spin and orbital angular momentum of the host element\cite{tanaka2008prb}. Such a trend has been found in experiments\cite{morota2011prb,liu2015apl}, suggesting that the intrinsic spin Hall effect is relevant to generating spin current in non-doped transition metals. 

Theoretical studies on the intrinsic spin Hall effect have proposed that the spin current is accompanied by a flow of orbital angular momentum, referred to as the orbital current, generated by the orbital Hall effect\cite{tanaka2008prb,kontani2009prl,go2018prl}. Orbital current consists of a flow of electrons with opposite orbital angular momentum moving in opposite directions. In contrast to the spin Hall conductivity, the orbital Hall conductivity is predicted to be independent of the size and sign of the spin orbit coupling ($\bm{L} \cdot \bm{S}$). Using tight binding and/or first-principles calculations, it has been shown that the 3$d$ transition metals, particularly V, Ti and Cr, possess one of the largest orbital Hall conductivities among the transition metals\cite{tanaka2008prb,jo2018prb,salemi2022prm}. 

Probing the orbital current remains a significant challenge in modern spintronics. A large number of studies have used bilayers that consist of a nonmagnetic metal (NM) and a ferromagnetic metal (FM) to assess the presence of orbital current. In almost all cases, the response of the FM layer magnetization to the current passed along the bilayer is used to study the size and direction of orbital current, if any, generated in the NM layer\cite{lee2021cphys,sala2022prr,bose2023prb,dutta2022prb,hayashi2023cphys}. There are a few difficulties in evaluating the orbital current in such systems. First, as many studies use the change in the magnetization direction with the current to extract information on the orbital current, it is essential that the orbital current exerts torque on the magnetization. However, it remains to be seen if the orbital current that enters the FM layer can exert torque on the FM layer magnetization in a way similar to that of spin current (see e.g. Refs.~\cite{go2020prr,lee2021ncomm}). Second, it is well known that the electronic and/or structural properties of the NM/FM interface can significantly influence the amplitude and the direction of spin current that flows into the FM layer\cite{stiles2002prb,haney2013prb,rojassanchez2014prl,zhang2015nphys}. The same issue also applies to orbital current. 
Finally, studies have shown that the presence of the FM layer can alter the spin transport properties within the NM layer via a proximity-like effect. For example, the spin diffusion length of the NM layer has been reported to differ depending on whether or not a FM layer is placed adjacent to the NM layer\cite{liu2011condmat}. Similar effects may also apply to the orbital current in the NM layer. For these reasons, it is preferable to probe the orbital current, if any, in a single NM layer.

The first observation of the spin Hall effect used the magneto-optical Kerr effect in a polar configuration to detect the amount of out-of-plane spin magnetic moments accumulated at the edges of a semiconductor\cite{kato2004science}. 
The Kerr rotation angle induced by the spin magnetic moments was a few microradians. 
Later on, a current-induced spin magnetic moment, polarized along the film plane and accumulated at the surface of a NM layer, was probed using the longitudinal Kerr effect\cite{vantErve2014apl,riego2016apl,su2017apl,stamm2017prl}. The associated Kerr signal in NMs was of the order of nanoradian. 
The difference in the Kerr signal between the former (semiconductors) and the latter (metals) is largely due to the difference in the spin diffusion length: the former has orders of magnitude larger spin diffusion length than the latter.
As the effect is not limited to detecting a spin magnetic moment, one may probe orbital magnetic moment at the surface/edges of nonmagnetic materials\cite{choi2023nature}.

Here we use the longitudinal magneto-optical Kerr effect to study current-induced spin and orbital magnetic moments in NM single-layer films. We compare the Kerr signal of Pt, known for its large spin Hall conductivity, and V, which is predicted to exhibit a large orbital Hall conductivity. 
A simple model is developed to analyze the experimental results. We show that the real and imaginary parts of the Kerr signal can be used to determine the spin and orbital Hall effects of the host material. From the analyses, we determine the spin and orbital diffusion length of Pt and V.

\section{Experimental results}
Films are deposited at ambient temperature using RF magnetron sputtering on thermally oxidized silicon substrates; the SiO$_2$ thickness is $\sim$100 nm.
The film structure is sub./$t$ NM/2 MgO/1 Ta (thickness in units of nanometer) where NM = Pt, V.
Optical lithography and Ar ion milling are used to pattern the films into a wire.
The width and length of the wire are 0.4 mm and 1.2 mm, respectively.
Electrodes made of 5 Ta/60 Cu/3 Pt are formed using conventional liftoff techniques.
X-ray diffraction is used to study the film structure. 
For V, we find the bcc (110) peak in the 20-nm-thick V thin film, suggesting that the film is composed of bcc-V.

\begin{figure}[t]
    \centering
    \includegraphics[width=0.9\linewidth]{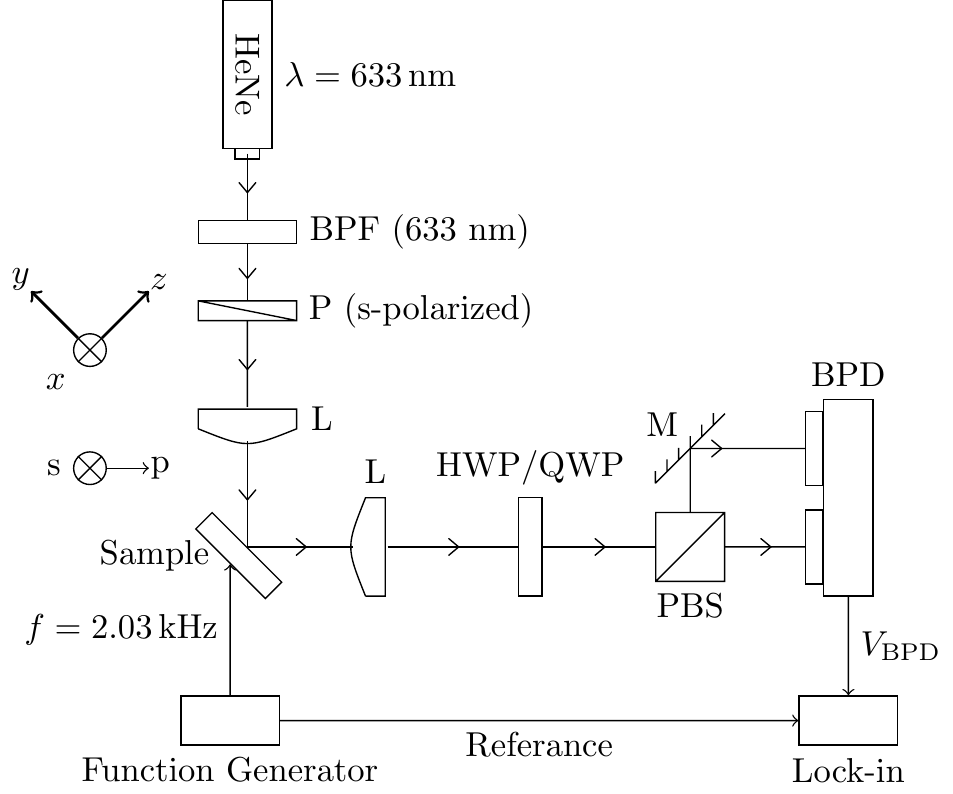}
    \caption{Schematic illustration of the optical setup and definition of the coordinate system. The light incidence of plane is the $yz$ plane, and the light is s-polarized: the polarization points along the $x$-axis. BPF: band-pass filter, P: polarizer, L: lens, HWP: half-wave plate, QWP: quarter-wave plate, PBS: polarized beam splitter, M: mirror, BPD: balanced photodetector.}
    \label{fig:setup}
\end{figure}

Schematic illustration of the experimental setup is shown in Fig.~\ref{fig:setup}.
An alternating current is applied to the sample along the $x$ direction from a function generator (Stanford Research Systems, DS360). The frequency of the current is fixed to 2030 Hz and the amplitude is varied. The sample is irradiated with a linearly polarized light from an oblique angle, $\sim$45 deg from the film normal. The light plane of incidence ($yz$ plane) is orthogonal to the current flow direction (along $x$). The reflected light is measured using a balanced photodetector, where the optical signal is converted to an electrical signal. The electrical output of the photodetector is fed into a lock-in amplifier. The frequency and phase of the lock-in amplifier are locked to those of the alternating current applied to the sample. A half-wave plate or a quarter-wave plate is inserted in between the sample and the photodetector. The half-wave plate and the quarter-wave plate are used to measure the real ($\theta_\mathrm{K}$, rotation angle) and imaginary ($\eta_\mathrm{K}$, ellipticity) parts of the Kerr signal, respectively. $\theta_\mathrm{K}$ and $\eta_\mathrm{K}$ are measured as a function of the current density $j$ applied to the sample. Such measurement is repeated $N$ times to improve the signal-to-noise ratio (typical $N = 100 - 1000$). 
The sample stage is controlled by a stepping motor-controlled system and we look for the sample position by scanning the laser across the sample and measuring the reflectivity. Once the center of the sample is identified, we fix the sample stage and perform the measurements. During the measurements, the laser beam is fixed at the center of the sample. Since the optical setup needs to be changed to measure $\theta_\mathrm{K}$ and $\eta_\mathrm{K}$, the two are measured separately.
All measurements are carried out at room temperature.

\begin{figure}[t]
    \centering
    \includegraphics[width=0.9\linewidth]{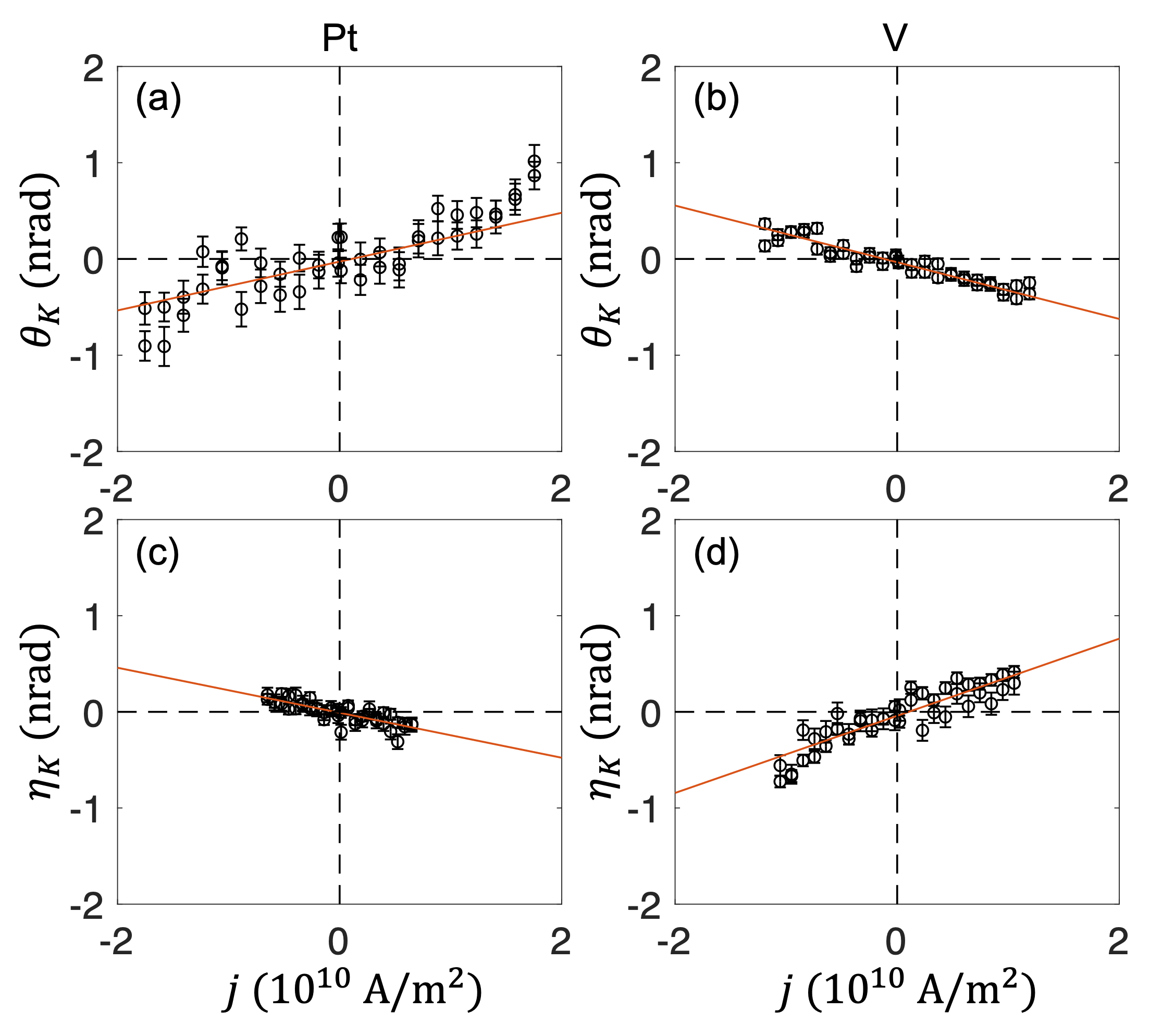}
    \caption{(a)-(d) Current density $j$ dependence of the real part $\theta_\mathrm{K}$ (a),(b) and the imaginary part $\eta_\mathrm{K}$ (c),(d) of the Kerr signal. Data are from 20 nm Pt (a),(c) and 20 nm V (b),(d). The error bars indicate standard deviation of the $N$ measurements conducted normalized by $\sqrt{N-1}$. Since $\theta_\mathrm{K}$ showed a nonlinear dependence on $j$ for Pt, we limited the measurement range of $\eta_\mathrm{K}$ to a smaller value.}
    \label{fig:kerr:j}
\end{figure}

The $j$ dependence of $\theta_\mathrm{K}$ and $\eta_\mathrm{K}$ are shown in Fig.~\ref{fig:kerr:j} for 20-nm-thick Pt and V films.
Positive and negative $j$ correspond to, respectively, in-phase and (180 deg) out-of-phase detection of the optical signal with respect to the current.
We find that both $\theta_\mathrm{K}$ and $\eta_\mathrm{K}$ linearly scales with $j$ at small $|j|$.
Data is fitted with a linear function to obtain the rate at which the Kerr signal varies with $j$.
Note that a slight non-linear $j$ dependence of the Kerr signal is found when $|j|$ exceeds $1 \times 10^{10}$ A/m$^2$: see, for example, Fig.~\ref{fig:kerr:j}(a).
The data fitting range is thus limited to below such value.
We also restricted the measurement range of $\eta_\mathrm{K}$ for Pt [Fig.~\ref{fig:kerr:j}(b)] to obtain a better linear fitting.
The nonlinear $j$ dependence is likely caused by current-induced heating and temperature-dependent optical properties\cite{footnote1}.
The effect is larger than that reported in Refs.~\cite{stamm2017prl,choi2023nature}, which may be due to the difference in the size of the wire used: the wire width is larger here by more than a factor of 10. As the power input to the sample scales the wire width, larger current-induced heating can take place in wider wires.
The large sample width, however, is beneficial for the optical measurements as the system is less susceptible to system vibration and laser spot drifting.

\begin{figure}[t]
    \centering
    \includegraphics[width=0.9\linewidth]{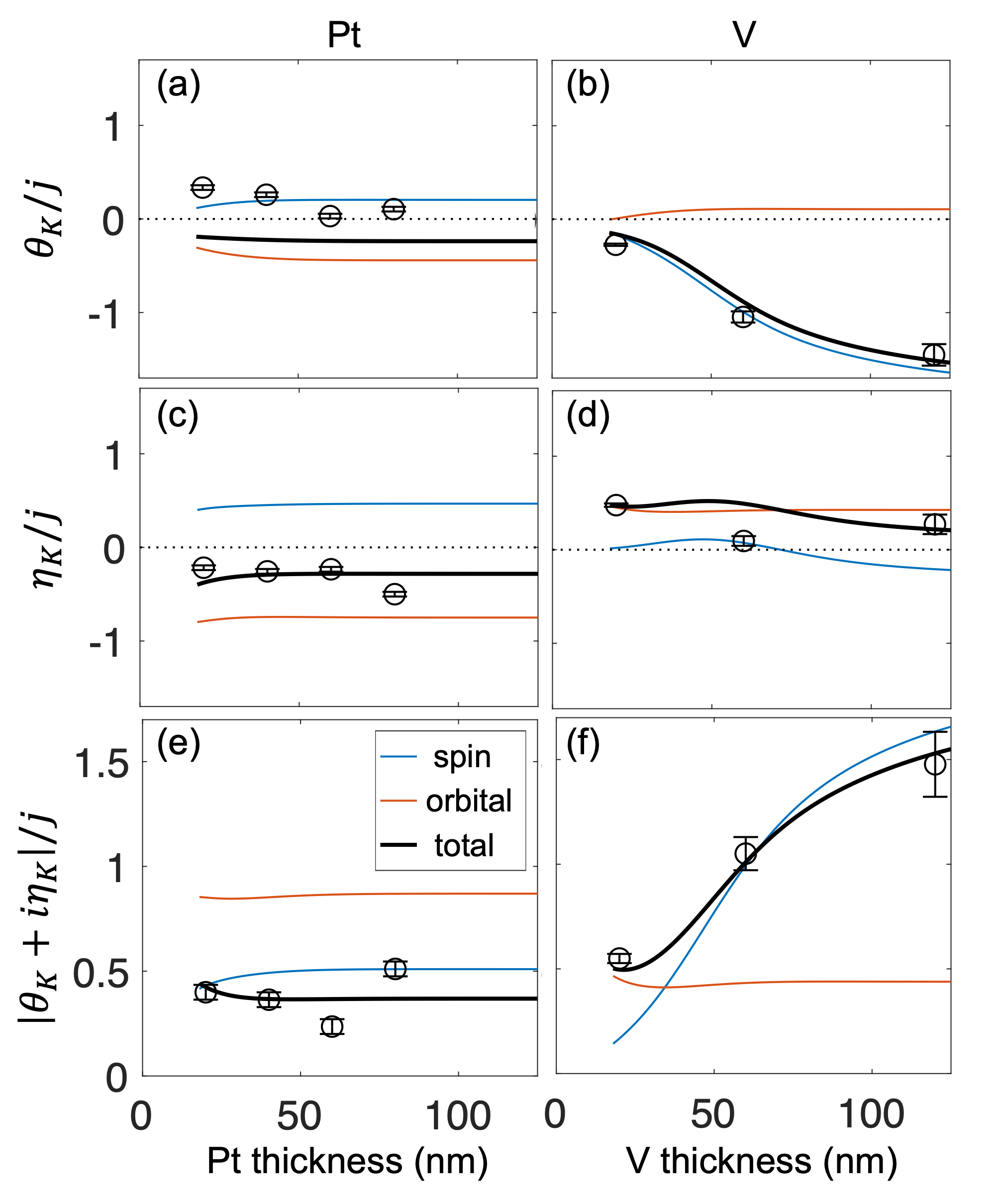}
    \caption{(a)-(f) The NM layer thickness dependence of the real part $\theta_\mathrm{K}$ (a),(b), the imaginary part $\eta_\mathrm{K}$ (c),(d) and the absolute value $\left| \theta_\mathrm{K} + i \eta_\mathrm{K} \right|$ (e),(f) of the Kerr signal divided by the current density $j$. All units are in 10$^{-10}$ nrad/(A/m$^2$).
    The circles show experimental data from Pt (a),(c),(e) and V (b),(d),(f). The error bars represent the fitting error of the $j$ dependence of the Kerr signal with a linear function. Blue, red, and black lines show the calculated Kerr signal with a contribution from the spin Hall effect, orbital Hall effect, and the sum of the two, respectively. Parameters used in the calculations are summarized in the second and fifth lines of Tables~\ref{table:para}. Calculation results for the thin NM layer limit are not shown for clarity: see Appendix~\ref{sec:app:model} for the details.
    }
    \label{fig:kerr:thickness}
\end{figure}

The film thickness dependence of $\theta_\mathrm{K} / j$ and $\eta_\mathrm{K} / j$ obtained from the linear fitting of the Kerr signal are presented in Fig.~\ref{fig:kerr:thickness}(a)-\ref{fig:kerr:thickness}(d) with the open symbols. For Pt, we find that both $\theta_\mathrm{K} / j$ and $\eta_\mathrm{K} / j$ show a relatively small dependence on the Pt layer thickness. The value of $\theta_\mathrm{K} / j$ is smaller than that reported in Refs.~\cite{stamm2017prl,choi2023nature}: the reason behind this is discussed later in this section. In contrast, $\theta_\mathrm{K} / j$ for V steadily increases with increasing V layer thickness ($\eta_\mathrm{K} / j$ shows a rather small variation). The absolute value of the Kerr signal $\left| \theta_\mathrm{K} + i \eta_\mathrm{K} \right| / j$ is plotted in Fig.~\ref{fig:kerr:thickness}(e) and \ref{fig:kerr:thickness}(f). As evident, the absolute Kerr signal is significantly larger for V than for Pt in the thick NM layer limit. 

To account for these experimental results, we model the current-induced magneto-optical Kerr effect assuming that spin and orbital magnetic moments, induced by spin and orbital Hall effects, respectively, accumulate at the surface of the film.
$\theta_\mathrm{K}$ and $\eta_\mathrm{K}$ are defined using the reflection coefficients $r_{ss}$ and $r_{ps}$:
\begin{equation}
\begin{aligned}
\label{eq:thetaKetaK}
\theta_\mathrm{K} = \mathrm{Re} \left[ \frac{r_{ps}}{r_{ss}} \right], \ \eta_\mathrm{K} = \mathrm{Im} \left[ \frac{r_{ps}}{r_{ss}} \right],
\end{aligned}
\end{equation}
where $r_{ss(ps)}$ represents the fraction of the $s$ ($p$) component of the reflected light when a sample is irradiated with an $s$-polarized light.
(The same convention applies when the sample is irradiated with a $p$-polarized light.)
$r_{ss}$ ($r_{ps}$) can be expressed using the diagonal ($\epsilon_{xx}$) and off-diagonal (e.g. $\epsilon_{zx}$) components of the permittivity tensor for each layer and geometrical parameters of the experimental setup (e.g. the light incident angle).  
For a sample made of nonmagnetic materials, the off-diagonal components are generally zero, which results in $r_{ps} = 0$.
Under the influence of spin and orbital Hall effects, current applied to the sample induces spin and/or orbital magnetic moments at the top and bottom surfaces of the film that cause one or more of the off-diagonal components to be non-zero.

We use the spin diffusion equation\cite{vanson1987prl} to obtain the profile of the spin/orbital magnetic moment.
Following the approach described in Appendix~\ref{sec:app:twocurrent}, 
$\epsilon_{zx}$ reads
\begin{equation}
\begin{gathered}
\label{eq:epsilon:xy}
\epsilon_{zx} = \epsilon_{zx,\mathrm{s}} + \epsilon_{zx,\mathrm{o}},\\
\epsilon_{zx,\mathrm{s(o)}} = - \frac{e l_\mathrm{s(o)} \rho_{xx}^2 \sigma_\mathrm{s(o)}(0)}{\epsilon_0 \omega } \frac{\sinh \left(\frac{t-2z}{2 l_\mathrm{s(o)}} \right)}{\cosh \left(\frac{t}{2 l_\mathrm{s(o)}} \right)} \frac{\partial \sigma_\mathrm{s(o)} (\omega)}{\partial E},
\end{gathered}
\end{equation}
where $\epsilon_0$ is the vacuum permittivity, $e$ is the electric charge, $\omega$ is the light angular frequency, and $\sigma_{xx}$, $\sigma_\mathrm{s(o)}(0)$, $l_\mathrm{s(o)}$ are the conductivity, dc spin (orbital) Hall conductivity and the spin (orbital) diffusion length, respectively, of the NM layer. 
$\sigma_\mathrm{s(o)} (\omega)$ is the NM layer ac spin (orbital) Hall conductivity at light angular frequency $\omega$ and $\frac{\partial \sigma_\mathrm{s(o)} (\omega)}{\partial E}$ is its energy derivative taken at the Fermi level.
$z = 0$ and $z = t$ correspond to bottom (facing the substrate) and top surfaces of the NM layer, respectively.

The Kerr signal is computed numerically.
Since $\epsilon_{zx}$ varies along the film thickness, the NM layer is divided into sub-layers with a thickness of $\delta t$ and a constant $\epsilon_{zx}$.
Here we set $\delta t = 0.1$ nm.
We neglect the effect of the capping layer (2 MgO/1 Ta) as it is transparent to visible light: the top Ta layer is oxidized due to exposure to air.
Following Zak \textit{et al}.\cite{zak1990jmmm}, we compute the $4 \times 4$ boundary $A_j$ and propagation $D_j$ matrices for each (sub)layer $j$. 
The reflection coefficients can be expressed as 
\begin{equation}
\begin{gathered}
\label{eq:M}
\begin{bmatrix}
r_{ss} &r_{sp}\\
r_{ps} &r_{pp}
\end{bmatrix}
= J G^{-1},\\
\begin{bmatrix}
G &H\\
J &K
\end{bmatrix}
= A_a^{-1} \prod_n \left( A_n D_n A_n^{-1} \right) \left( A_o D_o A_o^{-1} \right)A_s.
\end{gathered}
\end{equation}
where $G$, $H$, $J$ and $K$ are $2 \times 2$ matrices. $A_a$, $A_n$ ($D_n$), $A_o$ ($D_o$) and $A_s$ are the boundary (propagation) matrices for air, NM layer, SiO$_2$ layer and the Si substrate, respectively. Information on current-induced spin and orbital magnetic moments is encoded in $A_n$ and $D_n$ via $\epsilon_{zx}$ of the NM layer. 
Detailed forms of the matrices are described in the Appendix~\ref{sec:app:multilayers}.
Note that the light penetration depth, estimated from $\epsilon_{xx}$ (see Appendix~\ref{sec:app:ellips}), is of the order of a few tens of nanometers. 
Thus spin and orbital magnetic moments that accumulate at the top and bottom surfaces can contribute to the Kerr signal depending on the film thickness and the size of the spin and/or orbital diffusion lengths. 
Such an effect of the finite light penetration depth on the Kerr signal\cite{stamm2017prl,choi2023nature} is taken into account using this approach (see Appendix~\ref{sec:app:multilayers}).

\begin{figure}[t]
    \centering
    \includegraphics[width=0.9\linewidth]{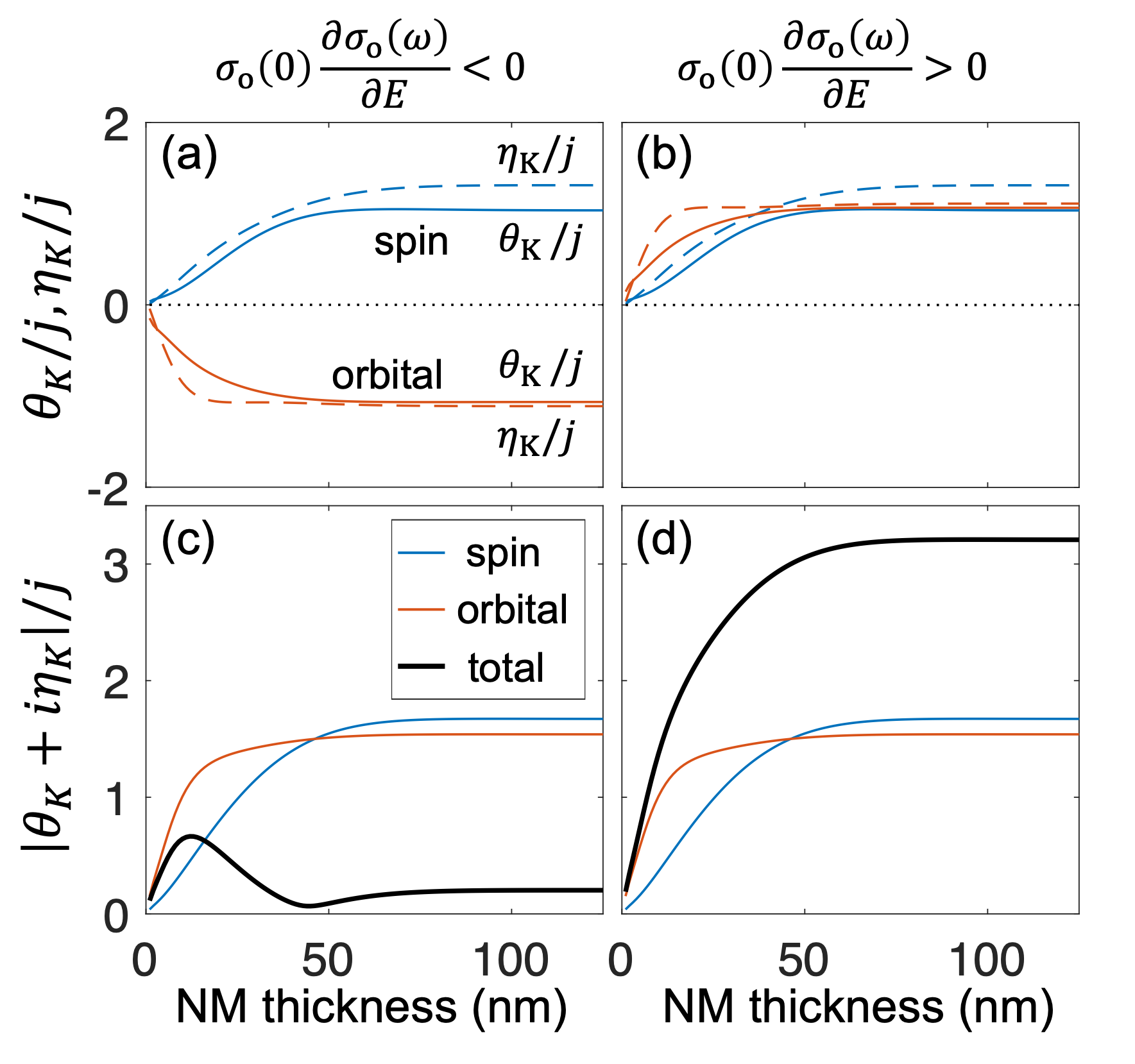}
    \caption{(a)-(d) The NM layer thickness dependence of the calculated $\theta_\mathrm{K}$ (solid line), $\eta_\mathrm{K} $ (dashed line) (a),(b) and the calculated $\left| \theta_\mathrm{K} + i \eta_\mathrm{K} \right| / j$ (c),(d). All units are in 10$^{-10}$ nrad/(A/m$^2$). Blue, red, and black lines show the contribution from the spin Hall effect, orbital Hall effect, and the sum of the two, respectively. Parameters used in the calculations are $l_\mathrm{s} = 10$ nm, $l_\mathrm{o} = 4$ nm, $\rho_{xx} = 20 \ \Omega \, $cm, refractive index $n = 2 + 4 i$ and the rest are shown in Table~\ref{table:para:gen}.}
    \label{fig:kerr:cal}
\end{figure}

We first show in Fig.~\ref{fig:kerr:cal} representative calculation results using parameter sets described in Table~\ref{table:para:gen} and in the caption to Fig.~\ref{fig:kerr:cal}.
The parameters are chosen such that only the sign of $\frac{\partial \sigma_\mathrm{o} (\omega)}{\partial E}$ is different for the two sets.
The values of parameters are chosen such that they resemble those of Pt.
To simplify the modeling, we neglect the SiO$_2$ layer of the substrate, which can influence the Kerr signal due to multiple reflections within SiO$_2$ when the NM layer is thin\cite{sumi2018scirep}.
The blue and red lines in Fig.~\ref{fig:kerr:cal} represent contributions from the spin and orbital Hall effects to the Kerr signal.
The solid and dashed lines in Fig.~\ref{fig:kerr:cal}(a) and \ref{fig:kerr:cal}(b) show $\theta_\mathrm{K}$ and $\eta_\mathrm{K}$, whereas the solid lines in Fig.~\ref{fig:kerr:cal}(c) and \ref{fig:kerr:cal}(d) represent the absolute value $|\theta_\mathrm{K} + i \eta_\mathrm{K}|$.
The thick black line in Fig.~\ref{fig:kerr:cal}(c) and \ref{fig:kerr:cal}(d) show the sum from the two (spin and orbit) contributions.
$\frac{\partial \sigma_\mathrm{o} (\omega)}{\partial E}$ is positive [negative] in Fig.~\ref{fig:kerr:cal}(a) and \ref{fig:kerr:cal}(c) [\ref{fig:kerr:cal}(b) and \ref{fig:kerr:cal}(d)].
\renewcommand{\arraystretch}{1.2}
\begin{table}[b]
 \caption{
Parameters used in the model calculations presented in Fig.~\ref{fig:kerr:cal}. The dc spin (orbital) Hall conductivity $\sigma_\mathrm{s(o)}(0)$ is in units of ($\Omega\, $cm)$^{-1}$, whereas the energy derivative of the ac spin (orbital) Hall conductivity $\frac{\partial \sigma_\mathrm{s(o)} (\omega)}{\partial E}$ is in units of ($\Omega\, $cm$\, $eV)$^{-1}$. 
}
 \label{table:para:gen}
 \centering
  \begin{tabular}{c c c c c}
   \hline \hline
    & $\sigma_\mathrm{s}(0)$ & $\frac{\partial \sigma_\mathrm{s} (\omega)}{\partial E}$ & $\sigma_\mathrm{o}(0)$ & $\frac{\partial \sigma_\mathrm{o} (\omega)}{\partial E}$\\
   \hline 
   $\mathrm{Fig.~\ref{fig:kerr:cal} (a)}$ & 1000 & $1000 + 1000 i $ & 2000 & $-2000 - 2000 i$\\
   $\mathrm{Fig.~\ref{fig:kerr:cal} (b)}$ & 1000 & $1000 + 1000 i $ & 2000 & $2000 + 2000 i$\\                          
         \hline
  \end{tabular}
\end{table}

As is evident, $|\theta_\mathrm{K} + i \eta_\mathrm{K}|$ in the thick NM limit is significantly larger for Fig.~\ref{fig:kerr:cal}(d) than Fig.~\ref{fig:kerr:cal}(c).
Equation~(\ref{eq:epsilon:xy}) shows that the Kerr signal scales with the sum of contributions from the spin and orbital Hall effects, each of which is proportional to the product of the dc spin or orbital Hall conductivity and the energy derivative of the ac spin or orbital Hall conductivity, i.e., Kerr signal $\propto \sigma_\mathrm{s} (0) \frac{\partial \sigma_\mathrm{s} (\omega)}{\partial E} + \sigma_\mathrm{o} (0) \frac{\partial \sigma_\mathrm{o} (\omega)}{\partial E}$.
If the sign of the product is the same (opposite) for spin and orbital Hall effects, both contributions add up (cancel out), which is the case for Fig.~\ref{fig:kerr:cal}(b) and \ref{fig:kerr:cal}(d) [Fig.~\ref{fig:kerr:cal}(a) and \ref{fig:kerr:cal}(c)].
As we show below, contributions from spin and orbital Hall effects cancel out for Pt whereas they add up for V.

\begin{table*}
 \caption{
Parameters for Pt and V used in the model calculations. The first and fourth lines show results from first-principles calculations. The second and fifth (third and sixth) lines present parameters for which the results are presented in Fig.~\ref{fig:kerr:thickness} (Fig.~\ref{fig:kerr:interference}). The light frequency $\omega$ corresponds to that of $\lambda = 633$ nm. $n$ is the refractive index of the film obtained from the ellipsometry measurements (see Appendix~\ref{sec:app:ellips}).}
 \label{table:para}
 \centering
   \vspace{3pt}
  \begin{tabular}{c c c c c c c c c c}
   \hline \hline
    Film & & $\sigma_\mathrm{s}(0)$ & $l_\mathrm{s}$ & $\frac{\partial \sigma_\mathrm{s} (\omega)}{\partial E}$ & $\sigma_\mathrm{o}(0)$ & $l_\mathrm{o}$ & $\frac{\partial \sigma_\mathrm{o} (\omega)}{\partial E}$ & $\rho_{xx}$ & $n$\\
    & & ($\Omega\, $cm)$^{-1}$ & nm & ($\Omega\, $cm$\, $eV)$^{-1}$ & ($\Omega\, $cm)$^{-1}$ & nm & ($\Omega\, $cm$\, $eV)$^{-1}$ & $\mu \Omega \, $cm &\\
   \hline 
   Pt & $\mathrm{DFT}$ & 2078 & n/a & $314 + 1540 i$ & 3240 & n/a & $-1620 - 5670 i$ & n/a & n/a\\
   & $\mathrm{Fig.~\ref{fig:kerr:thickness}}$ & 2078 & 4.0 & $314 + 1540 i$ & 3240 & 2.0 & $-1620 - 5670 i$ & 17 & $1.77 + 4.33 i$\\
   & $\mathrm{Fig.~\ref{fig:kerr:interference}}$ & 2078 & 4.0 & $1256 + 1540 i $ & 3240 & 2.0 & $-405 - 5670 i$ & 17 & $1.77 + 4.33 i$\\
         \hline
   V & $\mathrm{DFT}$  & -53 & n/a & $-16 - 101 i$ & 4180 & n/a & $1770 + 1280 i$ & n/a & n/a\\
   & $\mathrm{Fig.~\ref{fig:kerr:thickness}}$ & -154 & 40 & $160 - 290 i$ & 4180 & 0.7 & $1770 + 1280 i$ & 41 & $2.91 + 3.26 i$\\
   & $\mathrm{Fig.~\ref{fig:kerr:interference}}$ & -345 & 5.0 & $364 - 657 i$ & 196 & 35 & $-42 + 180 i$ & 41 & $2.91 + 3.26 i$\\      
         \hline
  \end{tabular}
\end{table*}

Next, we use the model to account for the experimental results.
Here we include the SiO$_2$ layer in the calculations.
$\epsilon_{xx}$ for the NM layer and the substrate is obtained from standard ellipsometry measurements (see Appendix~\ref{sec:app:ellips}).
The NM layer resistivity is measured with four-point probe technique. The dc and ac spin Hall conductivities are calculated using first-principles calculations: see Appendix~\ref{sec:app:dft} for details.
The obtained values are summarized in the first and fourth lines of Table~\ref{table:para}.
The spin ($l_\mathrm{s}$) and orbital ($l_\mathrm{o}$) diffusion length are used as fitting parameters to account for the data presented in Fig.~\ref{fig:kerr:thickness}, open symbols.
The parameters used in the calculations are summarized in the second and fifth lines of Table~\ref{table:para}.
For V, we slightly adjust the dc and ac spin Hall conductivities to fit the data. 
As the dc and ac spin Hall conductivity of V are small, we consider the accuracy of the calculations is lower than that of Pt and the orbital Hall conductivity of V. 
The value of the dc spin Hall conductivity $\sigma_\mathrm{s}(0) \sim -154$ $\left( \Omega \, \mathrm{cm} \right)^{-1}$ used in the model calculations is close to what has been found experimentally for bcc-V ($\sigma_\mathrm{s}(0) \sim -200$ $\left( \Omega \, \mathrm{cm} \right)^{-1} )$\cite{wang2017scirep}.

The model calculation results are shown by the solid lines in Fig.~\ref{fig:kerr:thickness}.
The blue and red lines represent contributions from the spin and orbital Hall effects to the Kerr signal, whereas the thick black line shows the sum of the two.
We find a relatively good agreement between the experimental results and the calculations.
First-principles calculations suggest that the signs of $\sigma_\mathrm{s} (0) \frac{\partial \sigma_\mathrm{s} (\omega)}{\partial E}$ and $\sigma_\mathrm{o} (0) \frac{\partial \sigma_\mathrm{o} (\omega)}{\partial E}$ are opposite for Pt while they are the same for V (see Table~\ref{table:para}).
From the model calculations, this suggests that the spin and orbital Hall contributions cancel out for Pt whereas they both add up for V.
(Strictly speaking, for V, the real part of the Kerr signal does not add up and causes a slight reduction in the overall signal.)
The cancellation explains why the observed Kerr signal is small for Pt compared to V.
This may also account for the difference in the Kerr signal found here and that reported in Refs.~\cite{stamm2017prl,choi2023nature}. 
Since the signal depends on a delicate balance between the spin and orbital Hall contributions, a slight change in either contribution can change the Kerr signal accordingly.
Although its influence may be minor, the light wavelength used here is also different from that in Refs.~\cite{stamm2017prl,choi2023nature}.

For Pt, the sign of $\theta_\mathrm{K} / j$ estimated from the model calculations does not agree with the experimental results. 
Such discrepancy can be resolved if the energy derivative of the ac spin and orbital Hall conductivities are changed from those obtained in the first-principles calculations by a factor of 4 and 1/4, respectively. 
In Fig.~\ref{fig:kerr:interference}(b), \ref{fig:kerr:interference}(d) and \ref{fig:kerr:interference}(f), the calculated Kerr signal is plotted with the adjusted parameters described in the third line of Table~\ref{table:para}.
As is evident, the experimental results are well reproduced with the adjusted parameter set.


For V, we find that the orbital diffusion length $l_\mathrm{o}$ must be smaller than a nanometer to account for the experimental results, which is in contrast to a previous study on a different transition metal (Ti)\cite{choi2023nature}. 
The small $l_\mathrm{o}$ is due to the large orbital Hall conductivity estimated from the first-principles calculations.
The experiments show that the Kerr signal increases with increasing V thickness and tends to saturate at $d \sim 100$ nm.
This suggests that $l_\mathrm{s}$ and/or $l_\mathrm{o}$ have a length scale of the order of a few tens of nm. 
Note that the spin/orbital diffusion length also sets the magnitude of the Kerr signal. 
Given the large orbital Hall conductivity of V estimated from first-principles calculations, $l_\mathrm{o}$ must be smaller than $\sim$1 nm in order to describe the Kerr signal found in the experiments.
This in turn requires a large $l_\mathrm{s}$ ($\sim$40 nm) with relatively small dc and ac spin Hall conductivities.
If we allow changes in the orbital Hall conductivity, the parameter set that describes the experimental results will vary.
We note that the overall thickness dependence of the Kerr signal cannot be accounted for with a single diffusion length: one needs short and long diffusion lengths to describe the experimental results.
Thus the presence of spin and orbital Hall effects is essential.
If we assume that the characteristics length scale is defined by the orbital Hall effect and $l_\mathrm{o}$ is equal to a few tens of nm, $\sigma_\mathrm{o}(0)$ needs to be reduced by more than a factor of 10 compared to what is suggested by the first principles calculations. 
The calculated results using such parameter set, described in the sixth line of Table~\ref{table:para}, are shown in Fig.~\ref{fig:kerr:interference}(h), \ref{fig:kerr:interference}(j), \ref{fig:kerr:interference}(l).
Clearly, the model calculations agree well with the experimental results.
Further study is required to identify the magnitude of $\sigma_\mathrm{o}(0)$ and $l_\mathrm{o}$ in an independent way.

\section{Conclusion}
In summary, we have studied the current-induced magneto-optical Kerr effect in Pt and V thin films. We find the Kerr signal of V is significantly larger than that of Pt when the film thickness is $\sim$100 nm. A model, developed to account for the experimental results, suggests that the Kerr signal is proportional to the sum of contributions from the spin and orbital Hall effects, each of which is proportional to the product of the dc spin or orbital Hall conductivity and the energy derivative of the ac spin or orbital Hall conductivity. We find that contributions from the spin and orbital Hall effects cancel out for Pt, whereas the two effects mostly add up for V, leading to a larger signal for the latter. The analyses suggest that the orbital diffusion length of V must be significantly smaller than the spin diffusion length, provided that the magnitude of the orbital Hall conductivity is close to what first-principles calculations predict. These results thus clarify the mechanism of the current-induced magneto-optical Kerr effect in metallic thin films.

\begin{acknowledgments}
This work was partly supported by JSPS KAKENHI (Grant Numbers 23H00176, 20J13860), JST CREST (JPMJCR19T3), MEXT Initiative to Establish Next-generation Novel Integrated Circuits Centers (X-NICS) and Cooperative Research Project Program of RIEC, Tohoku University.
\end{acknowledgments}

\section{Appendix}
\subsection{\label{sec:app:twocurrent} Current induced magneto-optical Kerr effect}
The aim of this section is to derive Eq.~(\ref{eq:epsilon:xy}).
In the following, superscript $\sigma$ represents either the spin or orbital magnetic moment. 

We start from the two current model to derive the relation between the off-diagonal conductivity and the spin and orbital Hall conductivity [Eq.~(\ref{eq:deltasigma:relations:SH})]
In the two-current model, transport properties are described using two channels: one with magnetic moment $+\sigma$ and the other with $-\sigma$.
The carrier density and the $ij$-component of the conductivity tensor for electrons with magnetic moment $\sigma$ are defined as $n^{\sigma}$ and $\sigma_{ij}^{\sigma}$, respectively.
With such a definition, the $ij$-component of the conductivity tensor and the spin (orbital) Hall conductivity are expressed as
\begin{equation}
\begin{aligned}
\label{eq:transcond:twocurrent}
\sigma_{ij} (\omega) &= \sigma_{ij}^{+\sigma} (\omega) + \sigma_{ij}^{-\sigma} (\omega),\\
\sigma_\mathrm{s(o)} (\omega)  &= \sigma_{ij}^{+\sigma} (\omega) - \sigma_{ij}^{-\sigma} (\omega).
\end{aligned}
\end{equation}
Hereafter, we focus on the off-diagonal component of the conductivity, i.e., the transverse conductivity $\sigma_{ij}(\omega)$ with $i \neq j$.
For non-magnetic materials without any external stimuli (e.g., electric field), we have $\sigma_{ij}^{+\sigma} (\omega) = -\sigma_{ij}^{-\sigma} (\omega)$, which yields $\sigma_{ij} (\omega) = 0$ and $\sigma_\mathrm{s(o)} (\omega)  = 2 \sigma_{ij}^{+\sigma} (\omega) = - 2 \sigma_{ij}^{-\sigma} (\omega)$.
Let us assume that a stimuli induces a change in $n^{\sigma}$, which in turn modulates $\sigma_{ij} (\omega)$ and $\sigma_\mathrm{s(o)} (\omega)$.
Under such a circumstance, the following relations hold:
\begin{equation}
\begin{aligned}
\label{eq:deltasigma}
\delta \sigma_{ij} (\omega) &= \frac{\partial \sigma_{ij}^{+\sigma} (\omega)}{\partial n^{+\sigma}} \delta n^{+\sigma} + \frac{\partial \sigma_{ij}^{-\sigma} (\omega)}{\partial n^{-\sigma}} \delta n^{-\sigma},\\
\delta \sigma_\mathrm{s(o)} (\omega)  &= \frac{\partial \sigma_{ij}^{+\sigma} (\omega)}{\partial n^{+\sigma}} \delta n^{+\sigma} - \frac{\partial \sigma_{ij}^{-\sigma} (\omega)}{\partial n^{-\sigma}} \delta n^{-\sigma},
\end{aligned}
\end{equation}
where $\delta n^{\sigma}$ is an infinitesimal change in $n^{\sigma}$ under the stimuli and $\delta \sigma_{ij} (\omega)$, $\delta \sigma_\mathrm{s(o)} (\omega)$ are the corresponding infinitesimal change in $\sigma_{ij} (\omega)$, $\sigma_\mathrm{s(o)} (\omega)$.
Next, we define $\delta n_c$ as an infinitesimal change in the average carrier density and $\delta n_\mathrm{s(o)}$ as an infinitesimal change in the density of spin (orbital) magnetic moment [referred to as spin (orbital) density hereafter] under the stimuli:
\begin{equation}
\begin{aligned}
\label{eq:ncns}
\delta n_c &\equiv \delta n^{+\sigma} + \delta n^{-\sigma},\\
\delta n_\mathrm{s(o)}^{\sigma} &\equiv \delta n^{+\sigma} - \delta n^{-\sigma}.
\end{aligned}
\end{equation}
Combining Eqs.~(\ref{eq:deltasigma}) and (\ref{eq:ncns}), we obtain
\begin{equation}
\begin{aligned}
\label{eq:deltasigma:ncns}
\delta \sigma_{ij} (\omega) =& \frac{1}{2} \left( \frac{\partial \sigma_{ij}^{+\sigma} (\omega)}{\partial n^{+\sigma}} + \frac{\partial \sigma_{ij}^{-\sigma} (\omega)}{\partial n^{-\sigma}} \right) \delta n_c\\
&+ \frac{1}{2} \left( \frac{\partial \sigma_{ij}^{+\sigma} (\omega)}{\partial n^{+\sigma}} - \frac{\partial \sigma_{ij}^{-\sigma} (\omega)}{\partial n^{-\sigma}} \right) \delta n_\mathrm{s(o)}^{\sigma},\\
\delta \sigma_\mathrm{s(o)} (\omega)  =& \frac{1}{2} \left( \frac{\partial \sigma_{ij}^{+\sigma} (\omega)}{\partial n^{+\sigma}} - \frac{\partial \sigma_{ij}^{-\sigma} (\omega)}{\partial n^{-\sigma}} \right) \delta n_c\\
&+ \frac{1}{2} \left( \frac{\partial \sigma_{ij}^{+\sigma} (\omega)}{\partial n^{+\sigma}} + \frac{\partial \sigma_{ij}^{-\sigma} (\omega)}{\partial n^{-\sigma}} \right) \delta n_\mathrm{s(o)}^{\sigma}.
\end{aligned}
\end{equation}
From Eq.~(\ref{eq:deltasigma:ncns}), the following relations emerge:
\begin{equation}
\begin{aligned}
\label{eq:deltasigma:relations}
\frac{\partial \sigma_{ij} (\omega)}{\partial n_c} = \frac{1}{2} \left( \frac{\partial \sigma_{ij}^{+\sigma} (\omega)}{\partial n^{+\sigma}} + \frac{\partial \sigma_{ij}^{-\sigma} (\omega)}{\partial n^{-\sigma}} \right) = \frac{\partial \sigma_\textrm{s(o)} (\omega)}{\partial n_\mathrm{s(o)}^\sigma},\\
\frac{\partial \sigma_{ij} (\omega)}{\partial n_\mathrm{s(o)}^\sigma} = \frac{1}{2} \left( \frac{\partial \sigma_{ij}^{+\sigma} (\omega)}{\partial n^{+\sigma}} - \frac{\partial \sigma_{ij}^{-\sigma} (\omega)}{\partial n^{-\sigma}} \right) = \frac{\partial \sigma_\textrm{s(o)} (\omega)}{\partial n_c}.
\end{aligned}
\end{equation}
The second relation is particularly important here: It states that the change in the transverse conductivity $\delta \sigma_{ij} (\omega)$ with the spin (orbital) density $\delta n_\mathrm{s(o)}^{\sigma}$ is equivalent to the change in the spin (orbital) Hall conductivity $\delta \sigma_\mathrm{s(o)} (\omega)$ with the average carrier density $\delta n_c$. 
By definition, we have 
\begin{equation}
\begin{aligned}
\label{eq:n2E}
\delta n_c = D_\mathrm{F} \delta E,
\end{aligned}
\end{equation}
where $\delta E$ is the associated infinitesimal change in energy and $D_\mathrm{F}$ is the density of states at the Fermi level.
Thus from Eqs.~(\ref{eq:deltasigma:relations}) and (\ref{eq:n2E}), we obtain
\begin{equation}
\begin{aligned}
\label{eq:deltasigma:relations:SH}
\frac{\partial \sigma_{ij} (\omega)}{\partial n_\mathrm{s(o)}^{\sigma}} = \frac{1}{D_\mathrm{F}} \frac{\partial \sigma_\textrm{s(o)} (\omega)}{\partial E}.
\end{aligned}
\end{equation}
Here we have assumed
\begin{equation}
\begin{aligned}
\label{eq:dos}
D_\mathrm{F}^{+\sigma} = D_\mathrm{F}^{-\sigma} \equiv \frac{D_\mathrm{F}}{2},
\end{aligned}
\end{equation}
since the system under consideration is a non-magnetic material.
Note that the quantity on the right-hand side of Eq.~(\ref{eq:deltasigma:relations:SH}) can be computed numerically using first-principles calculations.
Later in this section, we substitute Eq.~(\ref{eq:deltasigma:relations:SH}) into Eq.~(\ref{eq:deltatranscond}) and subsequently derive Eq.~(\ref{eq:epsilon:xy}).

Next we use the spin/orbital diffusion equation to derive the spin and orbital density induced by the spin and orbital Hall effects.
The chemical potential for carriers with magnetic moment along $\sigma$ is defined as 
\begin{equation}
\begin{aligned}
\label{eq:chempot}
\mu^{\sigma} (\bm{r}) = - e \phi (\bm{r}) + \frac{n^{\sigma} (\bm{r})}{D_\mathrm{F}/2}.
\end{aligned}
\end{equation}
where $\phi (\bm{r})$ is the electric potential. For the time being, we explicitly show the spacial dependence of relevant quantities.
We define the difference in the chemical potential with opposite magnetic moment as $\mu_\mathrm{s(o)}^\sigma (\bm{r})$:
\begin{equation}
\begin{aligned}
\label{eq:spinchempot}
\mu_\mathrm{s(o)}^{\sigma} (\bm{r}) \equiv \mu^{+\sigma} (\bm{r}) - \mu^{-\sigma} (\bm{r}).
\end{aligned}
\end{equation}
The spin (orbital) diffusion equation for carriers with magnetic moment $\sigma$ is given by
\begin{equation}
\begin{aligned}
\label{eq:sde}
\nabla^2 \mu_\mathrm{s(o)}^{\sigma} (\bm{r}) = \frac{\mu_\mathrm{s(o)}^{\sigma} (\bm{r})}{\l_\mathrm{s(o)}^2},
\end{aligned}
\end{equation}
where $l_\mathrm{s(o)}$ is the spin (orbital) diffusion length.
We define the spin (orbital) current $\bm{j}_\mathrm{s(o)}^{\sigma}$ as a vector that represents the flow of carriers with magnetic moment $\sigma$. 
With spatial variation of $\mu_\mathrm{s(o)}^{\sigma} (\bm{r})$ and under the influence of spin (orbital) Hall effect, $\bm{j}_\mathrm{s(o)}^{\sigma} (\bm{r})$ is given as:
\begin{equation}
\begin{aligned}
\label{eq:spincurrent}
\bm{j}_\mathrm{s(o)}^{\sigma}(\bm{r}) = - \frac{\sigma_{xx}}{2e} \nabla \mu^{\sigma}(\bm{r}) + \bm{j}_\mathrm{SH(OH)}^{\sigma},
\end{aligned}
\end{equation}
where $\bm{j}_\mathrm{SH(OH)}^{\sigma}$ is the spin (orbital) current induced by the spin (orbital) Hall effect and $\sigma_{xx}$ is the conductivity.

The system under consideration is a thin film with the film normal along the $z$ axis. Current is passed along the film plane (along $x$) and the spin (orbital) Hall effect induces spin (orbital) current that flows perpendicular to the current.
Here we consider spin (orbital) current that flows along the film normal (along $z$) since we measure, using the longitudinal magneto-optical Kerr effect, the spin (orbital) magnetic moment at the film surface.
We define $z= 0$ and $z=t$ as the bottom (interfacing the substrate) and top surfaces of the film, respectively.

The solution of the spin (orbital) diffusion equation~(\ref{eq:sde}) takes the form
\begin{equation}
\begin{aligned}
\label{eq:sde:sol}
\mu_\mathrm{s(o)}^\sigma(z) = A \cosh \left( \frac{z}{l_\mathrm{s(o)}} \right) + B \sinh \left( \frac{z}{l_\mathrm{s(o)}} \right),
\end{aligned}
\end{equation}
where $A$ and $B$ are constants.
The boundary conditions are given as
\begin{equation}
\begin{aligned}
\label{eq:sde:bc}
j_{\mathrm{s(o)},z}^{\sigma} (z = 0) = j_{\mathrm{s(o)},z}^{\sigma} (z = t) = 0.
\end{aligned}
\end{equation}
We obtain the following solution that satisfies the boundary conditions:
\begin{equation}
\begin{aligned}
\label{eq:mus}
\mu_\mathrm{s(o)}^{\sigma}(z) = - \frac{2 e l_\mathrm{s(o)} j_\mathrm{SH(OH)}^\sigma}{\sigma_{xx}} \frac{\sinh \left(\frac{t-2z}{2 l_\mathrm{s(o)}} \right)}{\cosh \left(\frac{t}{2 l_\mathrm{s(o)}} \right)}.
\end{aligned}
\end{equation}
Using the relation $n_\mathrm{s(o)}^{\sigma} = n^{+\sigma} - n^{-\sigma}$ and Eqs.~(\ref{eq:chempot}), (\ref{eq:spinchempot}), the current-induced spin (orbital) density is given by
\begin{equation}
\begin{aligned}
\label{eq:spinaccum}
n_\mathrm{s(o)}^{\sigma}(z) =  \frac{D_\mathrm{F}}{2} \mu_\mathrm{s(o)}^{\sigma}(z).
\end{aligned}
\end{equation}
Substituting Eq.~(\ref{eq:mus}) into Eq.~(\ref{eq:spinaccum}) and using the relation $j_\mathrm{SH(OH)}^{\sigma} = \frac{\sigma_\mathrm{s(o)}}{\sigma_{xx}} j_x = \rho_{xx} \sigma_\mathrm{s(o)} j_x$, the current-induced spin (orbital) density reads
\begin{equation}
\begin{aligned}
\label{eq:spinaccum:all}
n_\mathrm{s(o)}^{\sigma}(z) = - 2 e l_\mathrm{s(o)} \sigma_\mathrm{s(o)} \rho_{xx}^2 j_x \frac{D_\mathrm{F}}{2} \frac{\sinh \left(\frac{t-2z}{2 l_\mathrm{s(o)}} \right)}{\cosh \left(\frac{t}{2 l_\mathrm{s(o)}} \right)}.
\end{aligned}
\end{equation}
We assumed the Stoner enhancement factor\cite{stamm2017prl} equals 1 here.
Hereafter, we do not write the $z$-dependence of $n_\mathrm{s(o)}^{\sigma}$ explicitly. 


In the experiments, we use the longitudinal Kerr effect to probe the magnetic moments that are parallel to the $y$-axis. 
Thus the Kerr signal reflects current-induced changes of the $zx$-component of the conductivity tensor [i.e., $\sigma_{zx} (\omega)$].
Since $\sigma_{zx} (\omega)$ is modulated by current-induced spin (orbital) density $n_\mathrm{s(o)}^{\sigma}$, the following relation holds: 
\begin{equation}
\begin{aligned}
\label{eq:deltatranscond}
\frac{\sigma_{zx}(\omega)}{j_x} = \frac{\partial \sigma_{zx}(\omega)}{\partial n_\mathrm{s(o)}^\sigma} \frac{n_\mathrm{s(o)}^\sigma}{j_x}.
\end{aligned}
\end{equation}
Substituting Eqs.~(\ref{eq:deltasigma:relations:SH}) and (\ref{eq:spinaccum:all}) into Eq.~(\ref{eq:deltatranscond}), we obtain
\begin{equation}
\begin{aligned}
\label{eq:deltatranscond:spinHall}
\frac{\sigma_{zx}(\omega)}{j_x} &= \frac{1}{D_\mathrm{F}} \frac{\partial \sigma_\textrm{s(o)} (\omega)}{\partial E} \frac{n_\mathrm{s(o)}^{\sigma}}{j_x}\\
&= - \frac{\partial \sigma_\textrm{s(o)} (\omega)}{\partial E} e l_\mathrm{s(o)} \sigma_\mathrm{s(o)} \rho_{xx}^2 j_x \frac{\sinh \left(\frac{t-2z}{2 l_\mathrm{s(o)}} \right)}{\cosh \left(\frac{t}{2 l_\mathrm{s(o)}} \right)}.
\end{aligned}
\end{equation}
To obtain the relative permittivity tensor, we use the following relation:
\begin{equation}
\begin{aligned}
\label{eq:permittivity2conductivity}
\epsilon_{ij} &= \delta_{ij} + \frac{i}{\epsilon_0 \omega} \sigma_{ij}.
\end{aligned}
\end{equation}
The change in the $zx$-component of the permittivity tensor ($\epsilon_{zx}$) with current reads
\begin{equation}
\begin{aligned}
\label{eq:permittivity2conductivity:delta}
\epsilon_{zx} = \frac{i}{\epsilon_0 \omega} \sigma_{zx}.
\end{aligned}
\end{equation}
We substitute $\sigma_{zx}$ from Eq.~(\ref{eq:deltatranscond:spinHall}) into Eq.~(\ref{eq:permittivity2conductivity:delta}), which gives the relation shown in Eq.~(\ref{eq:epsilon:xy}).

\subsection{\label{sec:app:multilayers} Longitudinal Kerr effect in multilayers}
Given that the light penetration depth is finite (of the order of a few tens of nanometers) for the NM films and the spin and orbital density along the film normal (along $z$) is not constant, the contribution on the Kerr signal from layer at a given $z$ is different for different $z$.
In addition, multiple reflections at the interfaces (air/film, film/SiO$_2$, SiO$_2$/Si substrate) can influence the Kerr signal\cite{sumi2018scirep}.
These effects need to be taken into account to provide an accurate estimation of the Kerr signal. 
We thus follow the approach established by Zak \textit{et al}\cite{zak1990jmmm}.
The boundary matrix of layer $j$ is given by
\begin{widetext}
\begin{equation}
\begin{gathered}
\label{eq:boundarymatrix}
A_j = 
\begin{bmatrix}
1 &0 &1 &0\\
-\frac{i}{2} \frac{\sin\theta_j}{\cos\theta_j} \left( 1 + \cos^2\theta_j \right) Q_j &\cos\theta_j Q_j &\frac{i}{2} \frac{\sin\theta_j}{\cos\theta_j} \left( 1 + \cos^2\theta_j \right) Q_j &-\cos\theta_j\\
\frac{i}{2} \sin\theta_j Q_j n_j & -n_j &\frac{i}{2} \sin\theta_j Q_j n_j &-n_j\\
\cos\theta_j n_j &\frac{i}{2} \frac{\sin\theta_j}{\cos\theta_j} Q_j n_j &-\cos\theta_j n_j &-\frac{i}{2} \frac{\sin\theta_j}{\cos\theta_j} Q_j n_j
\end{bmatrix},
\end{gathered}
\end{equation}
\end{widetext}
where $n_j$ and $Q_j = i \frac{\epsilon_{zx}}{\epsilon_{xx}}$ are the refractive index and the normalized off-diagonal component of the permittivity tensor of layer $j$, respectively. The angle $\theta_j$ satisfies the Snell's law: $n_i \sin\theta_i = n_j \sin\theta_j$ (layer $i$ supersedes layer $j$ when the light enters the multilayer).
The propagation matrix for layer $j$ reads
\begin{equation}
\begin{gathered}
\label{eq:propagationmatrix}
D_j = 
\begin{bmatrix}
U_j\cos\beta_j &U_j\sin\beta_j &0 &0\\
-U_j\sin\beta_j &U_j\cos\beta_j &0 &0\\
0 &0 &U_j^{-1} \cos\beta_j &-U_j^{-1} \sin\beta_j\\
0 &0 &U_j^{-1} \sin\beta_j &U_j^{-1} \cos\beta_j\\
\end{bmatrix},
\end{gathered}
\end{equation}
where $U_j = \exp\left( -i \frac{2 \pi}{\lambda} n_j \cos\theta_j t_j \right)$, $\beta_j = \frac{\pi \sin\theta_j}{\lambda \cos\theta_j} n_j Q_j t_j$ and $t_j$ is the thickness of layer $j$.
Boundary matrices $A_j$ and $D_j$ are computed and substituted into Eq.~(\ref{eq:M}).
The matrix product is computed numerically to obtain the reflection coefficients $r_{ss}$ and $r_{ps}$. 

\subsection{\label{sec:app:dft} First-principles calculations}
\begin{figure}[ht!]
    \centering
    \includegraphics[width=0.9\linewidth]{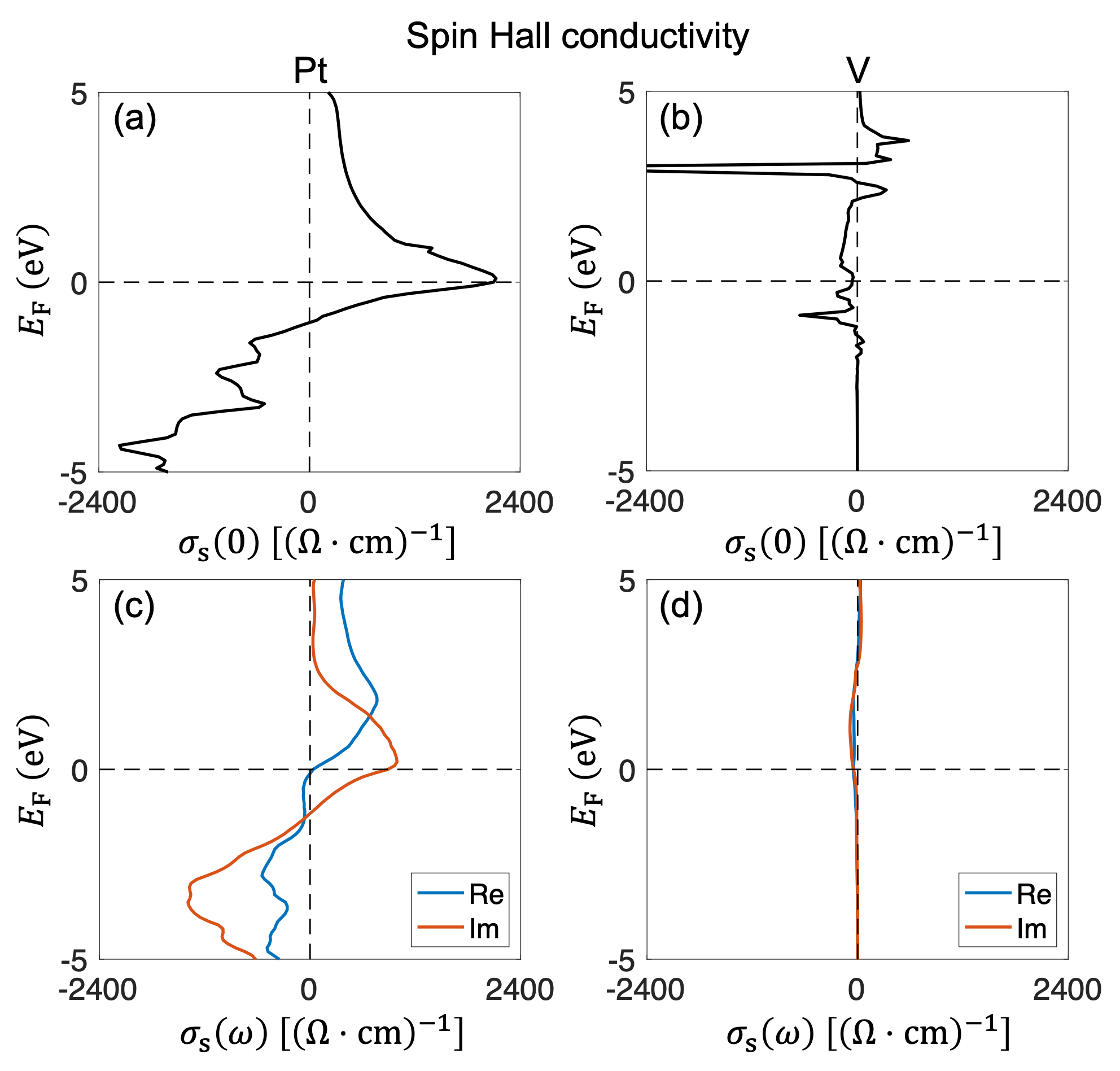}
    \includegraphics[width=0.9\linewidth]{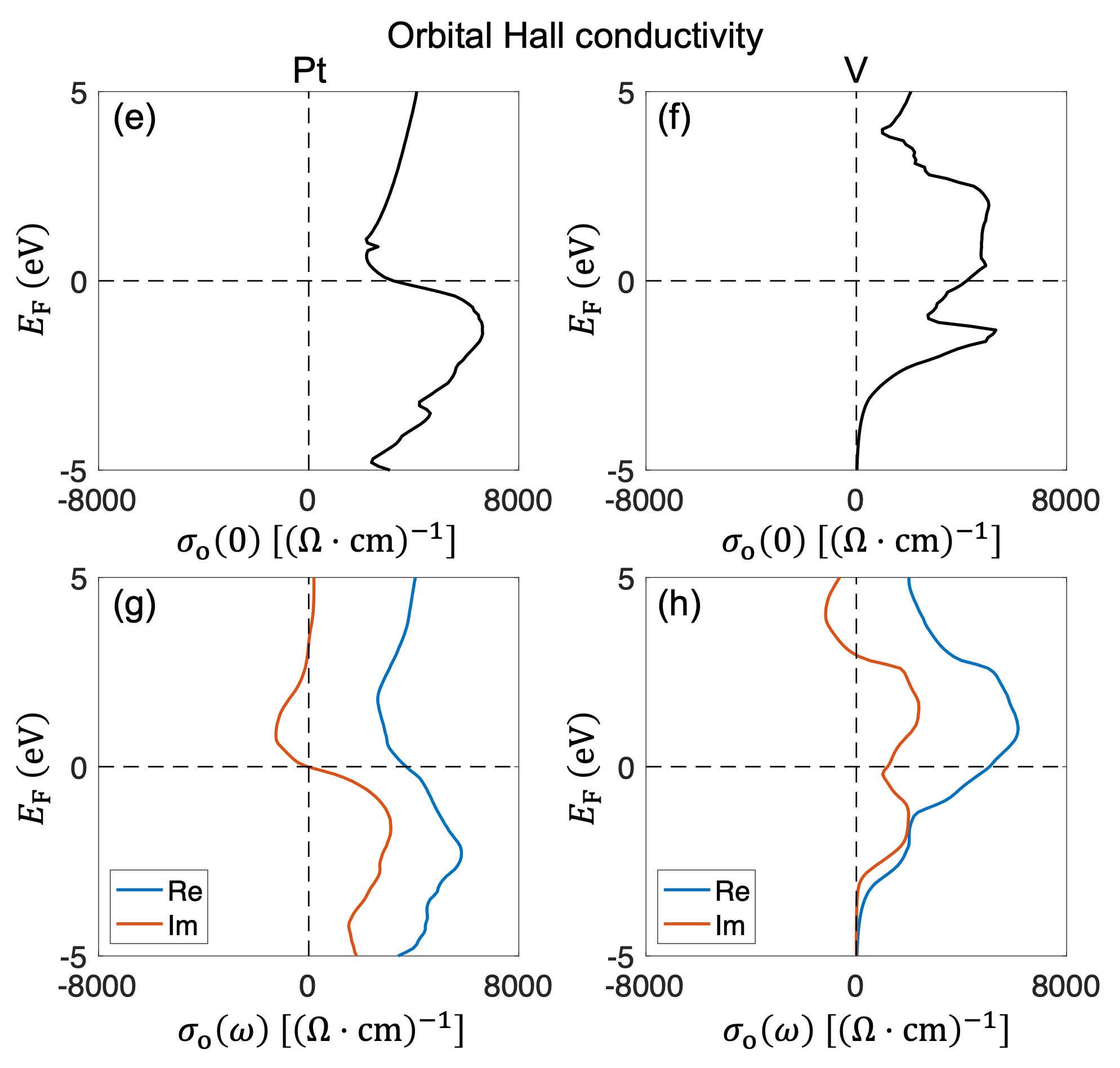}
    \caption{(a)-(d) First-principles calculations of the dc (a),(b) and ac (c),(d) spin Hall conductivities for Pt (a),(c) and V (b),(d). (e)-(h) First-principles calculations of the dc (e),(f) and ac (g),(h) orbital Hall conductivities for Pt (e),(g) and V (f),(h). The ac spin and orbital Hall conductivities are calculated using a light energy of 1.96 eV (the corresponding wavelength $\lambda$ is 633 nm). The blue and orange lines in (c),(d),(g),(h) represent the real and imaginary parts, respectively, of the ac spin/orbital Hall conductivity.}
    \label{fig:dft}
\end{figure}

First-principles calculations were performed using the all-electron full-potential linearized augmented plane wave (FLAPW) method\cite{nakamura2003prb} with generalized gradient approximation. 
The intrinsic dc spin (orbital) Hall conductivity $\sigma_{xy}^{s}(0)$ [$\sigma_{xy}^{o}(0)$] and the ac spin (orbital) Hall conductivity $\sigma_{xy}^{s}(\omega)$ [$\sigma_{xy}^{o}(\omega)$]\cite{wang1974prb,oppeneer1992prb,guo2008prl} are calculated using the linear-response Kubo formula\cite{guo2008prl,pradipto2018prb}.
For the latter, conductivities are calculated assuming that the system is under irradiation of light with energy of $\hbar \omega=1.96$ eV ($\lambda = 633$ nm), whereas the former corresponds to the static limit at $\omega=0$. 
The spin (orbital) current operator is defined as $j_{s,i}^{\sigma_j}=\frac{1}{2} \{s_j,v_i \}$ ($j_{o,i}^{\sigma_j}=\frac{1}{2} \{ l_j,v_i \}$), where $s_j$ ($l_j$) is the $j$-component of the spin (orbital) angular momentum operator, and $v_i$ is the $i$-component of the velocity operator. The curly brackets indicate anti-commutation: $\{A,B \}=AB+BA$. The relaxation time parameter ($1/\tau$) for the interband transition was set to 0.3 eV\cite{salemi2022prm}. The crystal structure is fcc for Pt and bcc for V (lattice constant, Pt: 0.392 nm, V: 0.303 nm). The Brillouin zone (BZ) integration was sampled by a special $k$=point mesh of $15 \times 15 \times 15$ for the self-consistent field and $51 \times 51 \times 51$ for the conductivities. The calculated dc and ac spin and orbital Hall conductivities for Pt and V are shown in Fig.~\ref{fig:dft}. 

\subsection{\label{sec:app:ellips} Determination of relative permittivity}
\begin{figure}[h!]
    \centering
    \includegraphics[width=0.9\linewidth]{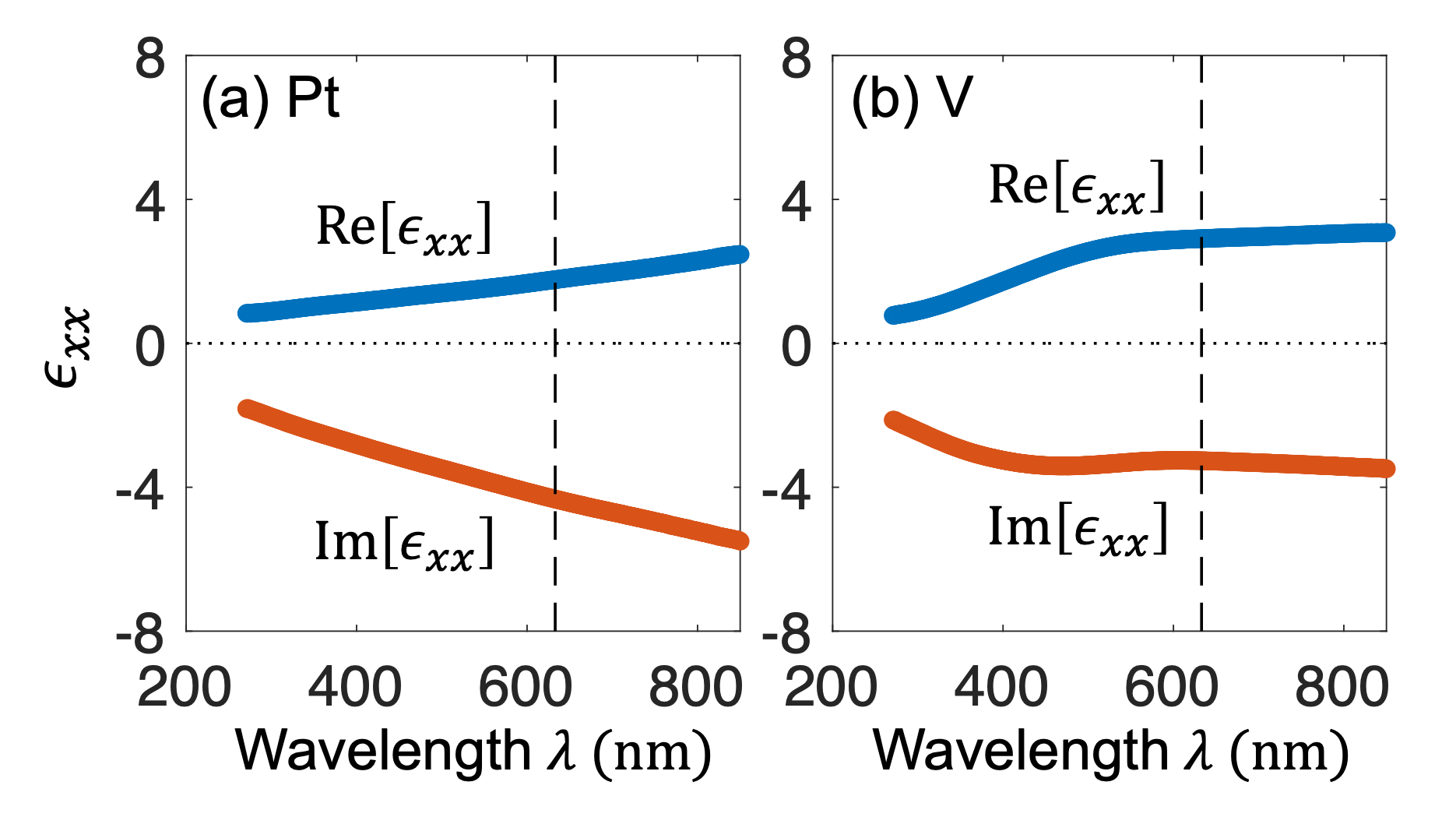}
    \caption{The light wavelength dependence of the relative permittivity $\epsilon_{xx}$ of the NM layer. Data from Pt (a) and V (b) thin films.}
    \label{fig:ellipsometry}
\end{figure}

Standard ellipsometry is used to determine the diagonal component of the permittivity tensor, i.e., the relative permittivity $\epsilon_{xx}$, of the materials under study. Here we use sub./80 Pt and sub./120 V/2 MgO/1 Ta to obtain $\epsilon_{xx}$ of the NM layer. 
The measured $\epsilon_{xx}$ of the films are shown in Fig.~\ref{fig:ellipsometry}.
At the light wavelength used in the experiments ($\lambda = 633$ nm), the light penetration depth [$1 / \left( 2 \pi \lambda \mathrm{Im}[n] \right) $] of the films is estimated to be 23 nm for Pt and 31 nm for V.
The refractive index of the Si substrate is measured using the same technique, from which we obtain $3.7 + 0.12 i$ at the light wavelength used in the experiments ($\lambda = 633$ nm). The refractive index ($n$) of SiO$_2$ and air are set to 1.5 and 1.0, respectively.

\subsection{\label{sec:app:model} Model calculations}
\subsubsection{\label{sec:app:model:multiplerefl} Effect of multiple reflections}
\begin{figure*}[h!]
    \centering
    \includegraphics[width=0.9\linewidth]{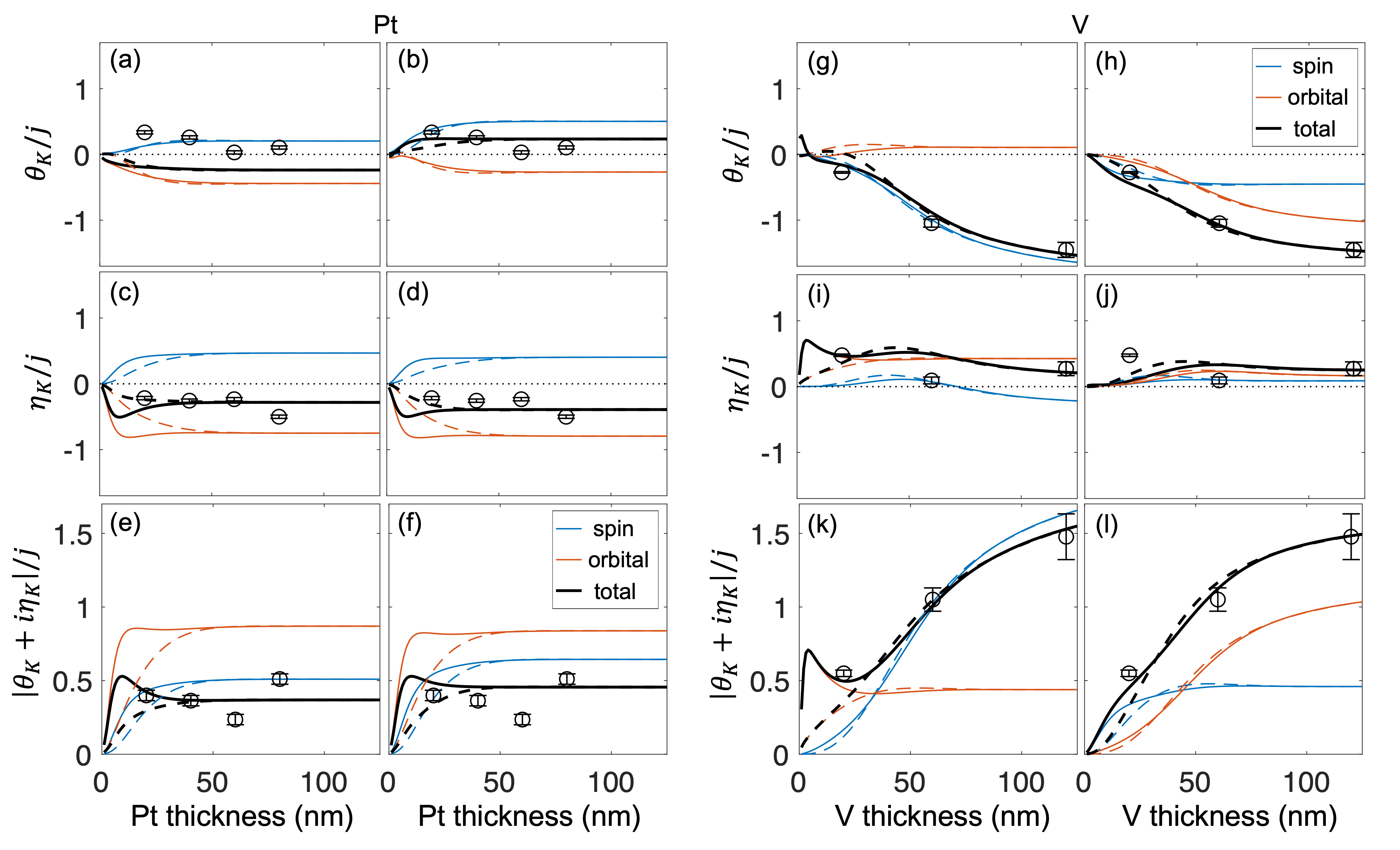}
    \caption{
    The NM layer thickness dependence of the real part $\theta_\mathrm{K}$ (a),(b),(g),(h), the imaginary part $\eta_\mathrm{K}$ (c),(d),(i),(j) and the absolute value $\left| \theta_\mathrm{K} + i \eta_\mathrm{K} \right|$ (e),(f),(k),(l) of the Kerr signal divided by the current density $j$. All units are in 10$^{-10}$ nrad/(A/m$^2$). Blue, red, and black lines show calculated Kerr signal with a contribution from the spin Hall effect, the orbital Hall effect, and the sum of the two, respectively. The solid lines show calculation results when a 100-nm-thick SiO$_2$ layer is placed in between the Si substrate and the NM layer. The dashed lines display results without the SiO$_2$ layer. Parameters used in the calculations are summarized in the second line (a),(c),(e), third line (b,) (d), (f), fifth line (g),(i),(k) and sixth line (h),(j),(l) of Table~\ref{table:para}. Note that the model calculations presented here do not necessarily show the best fit to experimental results; they are used to show how the Kerr signal changes as we vary the parameters.}
    \label{fig:kerr:interference}
\end{figure*}

In the experiments, we used Si substrates coated with a 100-nm-thick SiO$_2$ layer to avoid current shunting into the substrate. Multiple reflections within the SiO$_2$ layer can occur, which may influence the Kerr signal\cite{sumi2018scirep}. The effect of multiple reflections within the SiO$_2$ layer is larger when the NM layer thickness is smaller. To study such an effect, the Kerr signal is calculated with and without the SiO$_2$ layer. The results are presented in Fig.~\ref{fig:kerr:interference}(a)-\ref{fig:kerr:interference}(f) for Pt and Fig.~\ref{fig:kerr:interference}(g)-\ref{fig:kerr:interference}(l) for V. 
The solid and dashed lines show calculation results with and without the SiO$_2$ layer. As is evident, a clear difference is found when the NM layer thickness is smaller than 30 - 40 nm, which corresponds to the light penetration depth.

\subsubsection{\label{sec:app:model:fitpar} Fitting parameters: \textrm{Pt}}
With the parameter set described in the second line of Table~\ref{table:para}, we cannot account for the sign of $\theta_\mathrm{K} / j$ found in the experiments using the model calculations. This is because the dc and ac spin/orbital Hall conductivities are fixed to the values obtained from the first-principles calculations. 
If this restriction is lifted, one can reproduce the experimental results. Figure~\ref{fig:kerr:interference}(b),\ref{fig:kerr:interference}(d) and \ref{fig:kerr:interference}(f) show the calculated Kerr signal when the energy derivative of the real component of the ac spin and orbital Hall conductivities are changed. 
The other parameters are not changed: see the third line of Table~\ref{table:para}.
The calculated results show good agreement with the experimental results.
These results show the importance of determining the ac spin and orbital Hall conductivities in an accurate way. 

\subsubsection{\label{sec:app:model:fitpar} Fitting parameters: \textrm{V}}
For V, we find that the orbital diffusion length $l_\mathrm{o}$ must be smaller than a nanometer if we use the values of dc and ac orbital Hall conductivities obtained from the first-principles calculations. 
Such small $l_\mathrm{o}$ is in contrast to what has been found in Ti\cite{choi2023nature}.
Here we examine to what extent the dc and ac orbital Hall conductivities need to be varied to describe the experimental results under the constraint of a large orbital diffusion length reported in previous studies\cite{sala2022prr,choi2023nature}.
For this purpose, we fix $l_\mathrm{o} = 35$ nm.
Values of the dc and ac orbital Hall conductivities are then defined by the maximum Kerr signal of the thickest V film. We multiply a common factor to $\sigma_\mathrm{o}(0)$ and $\sigma_\mathrm{o}(\omega)$ obtained from the first-principles calculations. Parameters related to the spin Hall effect are set such that the Kerr signal of the thinner films can be accounted for. The calculation results are shown in Fig.~\ref{fig:kerr:interference}(h), \ref{fig:kerr:interference}(j) and \ref{fig:kerr:interference}(l) and the parameters used are summarized in the sixth line of Table~\ref{table:para}. As is evident, the calculation results agree well with the experimental results. Here, $\sigma_\mathrm{o}(0)$ and $\sigma_\mathrm{o}(\omega)$ are roughly 20 times smaller than those estimated from the first-principles calculations. 

\bibliography{ref_100223}

\begin{thebibliography}{43}
\bibitem [{foo()}]{footnote1}%
  \BibitemOpen
  \href@noop {} {}\bibinfo {note} {We comment on how current induced thermal
  effect manifest itself in the Kerr signal. Let us assume that the ac current
  applied to the sample varies with $\sin (\omega t)$. The spin and orbital
  accumulation change with $a \sin (\omega t)$, while any Joule heating induced
  effect will vary with $b \sin^2 (\omega t)$. Here the coefficient $a$ scales
  with the current density $j$ whereas $b$ is proportional to $j^2$. If any of
  the material parameters that determine the Kerr signal (e.g. refractive
  index) is temperature dependent, the Kerr signal will scale with the product
  of the two, i.e. $a \sin (\omega t) b \sin^2 (\omega t)$. Using the trigonal
  relation $\sin^3 (\omega t) = \frac{1}{4} \left( 3 \sin (\omega t) - \sin (3
  \omega t) \right)$, we find a Kerr signal that is proportional to $ab \sim
  j^3$.}\BibitemShut {Stop}%
\end{thebibliography}

\end{document}